\documentstyle[12pt,cite,axodraw]{article}
\input{epsf.sty}
\def\a{\alpha}
\def\b{\beta}

\def\d{\delta}
\def\e{\epsilon}

\def\g{\gamma}

\def\m{\mu}
\def\n{\nu}

\def\p{\pi}

\def\s{\sigma}
\def\t{\tau}

\def\F{\Phi}
\def\G{\Gamma}

\def\L{\Lambda}

\def\Q{\Theta}

\def\one{1\!\! 1}

\def\vf{\varphi}

\def\cm{{\cal M}}

\def\co{{\cal O}}

\def\10{{\bf 10}}

\def\5b{{\bf 5^*}}
\def\1{{\bf 1}}

\def\bo{{\raise.15ex\hbox{\large$\Box$}}}               % D'Alembertian
\def\pa{\partial}                                       % curly d
\def\pr{\prod}                                          % product
\def\face{{\raise.2ex\hbox{$\displaystyle \bigodot$}\mskip-2.2mu \llap {$\ddot
        \smile$}}}                                      % happy face
\def\dg{\dagger}                                     % hermitian conjugate
\def\wt#1{\widetilde{#1}}                    % big tilde
\def\Bar#1{\overline{#1}}                       % big bar
\def\VEV#1{\left\langle #1\right\rangle}        % < >
\def\leftrightarrowfill{$\mathsurround=0pt \mathord\leftarrow \mkern-6mu
        \cleaders\hbox{$\mkern-2mu \mathord- \mkern-2mu$}\hfill
        \mkern-6mu \mathord\rightarrow$}       % <--> double differential
\def\dvec#1{\vbox{\ialign{##\crcr
        \leftrightarrowfill\crcr\noalign{\kern-1pt\nointerlineskip}
        $\hfil\displaystyle{#1}\hfil$\crcr}}}           % <--> accent
\def\beq{\begin{equation}}
\def\eeq{\end{equation}}
\def\bea{\begin{eqnarray}}
\def\eea{\end{eqnarray}}
\def\NO{\nonumber}

\def\pl#1#2#3{Phys.~Lett.~{\bf B {#1}} (19{#2}) #3}
\def\np#1#2#3{Nucl.~Phys.~{\bf B {#1}} (19{#2}) #3}
\def\prl#1#2#3{Phys.~Rev.~Lett.~{\bf #1} (19{#2}) #3}
\def\pr#1#2#3{Phys.~Rev.~{\bf D {#1}} (19{#2}) #3}

\def\prep#1#2#3{Phys.~Rep.~{\bf {#1}C} (19{#2}) #3}

\begin{document}
\date{\mbox{ }}

\title{\Large 
{\normalsize     
DESY 99-021\hfill\mbox{}\\
April 1999\hfill\mbox{}\\}
\vspace{2cm}
\bf NEUTRINO MIXING AND THE PATTERN\\ OF SUPERSYMMETRY BREAKING\\[8mm]}
%
%\vspace{2cm} 
\author{Wilfried~Buchm\"uller, David~Delepine, Francesco~Vissani\\
{\it Deutsches Elektronen-Synchrotron DESY, Hamburg, Germany}}
\maketitle

\thispagestyle{empty}

\vspace{1cm}
\begin{abstract}
\noindent
We study the implications of a large $\n_\m$-$\n_\t$ mixing angle on
lepton flavour violating radiative transitions in supersymmetric extensions
of the standard model. The transition rates are calculated to leading order
in $\e$, the parameter which characterizes the flavour mixing. The uncertainty
of the predicted rates is discussed in detail. For models with modular 
invariance the branching ratio $BR(\m \rightarrow e \g)$ mostly exceeds the
experimental upper limit. In models with radiatively induced flavour mixing 
the predicted range  includes the 
upper limit, if the Yukawa couplings in the
lepton sector are large, as favoured by Yukawa coupling unification.
\end{abstract}

\newpage

In connection with the recently reported atmospheric neutrino anomaly 
\cite{atm98} the possibility of neutrino oscillations associated with
a large $\n_\m$-$\n_\t$ mixing angle has received wide attention. The
smallness of the corresponding neutrino masses can be accounted for 
by the seesaw mechanism \cite{yan79}, which leads to the prediction of heavy
Majorana neutrinos with masses close to the unification scale $\L_{GUT}$. 

A large $\n_\m$-$\n_\t$ mixing angle as well as the mass hierarchies of
quarks and charged leptons can be naturally explained by the Frogatt-Nielsen
mechanism based on a U(1)$_F$ family symmetry \cite{fro79} together with a 
nonparallel family structure of chiral charges \cite{sat98,ram98,bij87}. 
Depending on the family symmetry, such models can also explain the magnitude
of the observed baryon asymmetry \cite{buc99}. The expected phenomenology
of neutrino oscillations depends on details of the model \cite{vis98,irg98}.  

The large hierarchy between the electroweak scale and the unification scale, 
and now also the mass scale of the heavy Majorana neutrinos, motivates 
supersymmetric extensions of the standard model \cite{nil84}. This is further
supported by the observed unification of gauge couplings. The least understood
aspect of the supersymmetric standard model is the mechanism of supersymmetry 
breaking and the corresponding pattern of soft supersymmetry breaking
masses and couplings.

It is well known that constraints from rare processes severely restrict
the allowed pattern of supersymmetry breaking \cite{nil84}. In this paper 
we therefore study lepton flavour changing radiative transition 
\cite{ell82}. In the standard scenario with universal soft
breaking terms at the GUT scale, radiative corrections induce flavour
mixing at the electroweak scale. These effects can be important in the
case of large Yukawa couplings \cite{bar94,his98}. Following \cite{leo98} 
we shall contrast these minimal models with the interesting class of models 
possessing modular invariance \cite{iba92}.

In this paper we shall restrict our discussion to one particular example
with SU(5)$\otimes$U(1)$_F$ symmetry \cite{sat98,buc99}. However, the
results will be presented in such a form that they can easily be applied
to other examples of lepton mass matrices \cite{ros99}. We shall also
address the uncertainty of the predicted lepton flavour changing transition
rates.  

We consider the leptonic sector of the supersymmetric standard model with 
right-handed neutrinos, which is described by the superpotential
\beq\label{superp}
W = h_{e ij} \hat E^c_i \hat L_j \hat H_1 
    +  h_{\n ij} \hat N^c_i \hat L_j \hat H_2
    + \m \hat H_1 \hat H_2  +  \frac{1}{2} h_{r ij} \hat N^c_i \hat N^c_j \hat R\;.
\eeq
Here $i,j=1\ldots 3$ are generation indices, and the superfields $\hat E^c$,
$\hat L=(\hat N,\hat E)$, $\hat N^c$ contain the leptons $e_R^c$, $(\n_L,e_L)$,
$\n_R^c$, respectively. The expectation values of the Higgs multiplets $H_1$ 
and $H_2$ generate ordinary Dirac masses of quarks and leptons, and the 
expectation value of the singlet Higgs field $R$ yields the Majorana mass 
matrix of the right-handed neutrinos. 

In the following discussion the scalar masses will play a crucial role. They
are determined by the superpotential and the soft breaking terms,
\beq\label{soft}
{\cal{L}}_{soft} = -{\wt m}^2_{l ij} L^\dg_i L_j - {\wt m}^{2\dagger}_{e ij} E^{c\dg}_iE^c_j
           + A_{e ij} E^c_i L_j H_1 + c.c. + \ldots\;,
\eeq 
where $L=(N_L,E_L)$ and $E^c\equiv E_R^*$ denote the scalar partners of
$(\n_L,e_L)$ and $e_R^c$, respectively. Using the seesaw mechanism to explain 
the smallness of neutrino masses,
we assume that the right-handed neutrino masses $M_i$ are much larger than
the Fermi scale $v$. One then easily verifies that all mixing effects
on light scalar masses caused by the right-handed neutrinos and their
scalar partners are suppressed by $\co(v/M_i)$, and therefore negligible.

The mass terms of the light scalar leptons are given by 
\beq
{\cal{L}}_M = - E^\dg {\wt M_e}^2 E - N_L^\dg {\wt m_l}^2 N_L\;,
\eeq
where $\wt M_e^2$ is the mass matrix of the charged scalar fields 
$E=(E_L,E_R)$,
\beq\label{scalarm}
{\wt M_e}^2\ \equiv\ 
\left(\begin{array}{ll}
    {\wt M_L}^2 & {\wt M_{LR}}^2 \\[1ex]
     {\wt M_{RL}}^2 & {\wt M_R}^2  \\[1ex]
    \end{array}\right)\
=\ \left(\begin{array}{ll}
    {\wt m_l}^2 + v_1^2 h_e^\dg h_e  & v_1 A_e^\dg + \m v_2 h_e^\dg \\[1ex]
     v_1 A_e + \m v_2 h_e & {\wt m_e}^2 + v_1^2 h_e h_e^\dg \\[1ex]
    \end{array}\right)\;.
\eeq

According to the Frogatt-Nielsen mechanism \cite{fro79} the hierarchies
among the various Yukawa couplings are related to a spontaneously broken 
U(1)$_F$ generation symmetry. The Yukawa couplings arise from 
non-renormalizable interactions after a gauge singlet field $\F$ acquires
a vacuum expectation value,
\beq
h_{ij} = g_{ij} \left({\VEV\F\over \L}\right)^{Q_i + Q_j}\;.
\eeq
Here $g_{ij}$ are couplings $\co(1)$ and $Q_i$ are the U(1) charges of the
various superfields with $Q_{\F}=-1$. The interaction scale $\L$ is
expected to be very large, $\L > \L_{GUT}$, and the phenomenology of
quark and lepton mass matrices can be explained assuming 
\beq
\left({\VEV\F\over\L}\right)^2 \equiv \e^2 \simeq 
\left({m_\m \over m_\t}\right)^2 \simeq {1\over 300}\;. 
\eeq

\begin{table}[b]
\begin{center}
\begin{tabular}{c|ccccccccc}\hline \hline
$\hat\F_i$ & $\hat E^c_{3}$ & $\hat E^c_{2}$ & $\hat E^c_{1}$ & $\hat L_3$ & 
$\hat L_2$ & $\hat L_1$ &
$\hat N^c_{3}$ & $\hat N^c_{2}$ & $\hat N^c_{1}$ \\ \hline
$Q_i$ & 0 & 1 & 2 & $a$ & $a$ & $a+1$ & 0 & $1-a$ & $2-a$ \\ \hline\hline
\end{tabular}
\medskip
\caption[dum]{{\it Chiral charges for lepton superfields; a=0 or 1} \cite{buc99}.}
\end{center}
\end{table} 

The special feature of the two sets of charges $Q_i$ in table~1 is the
non-parallel family structure. The assignment of the same charge to the
lepton doublets of the second and third generation leads to a neutrino
mass matrix of the form\cite{sat98,ram98}, 
\beq\label{matrix}
m_{\n_{ij}} \sim \e^a \left(\begin{array}{ccc}
    \e^2  & \e  & \e \\[1ex]
    \e  &  1  & 1 \\[1ex]
    \e  &  1  & 1 
    \end{array}\right) {v^2\over \VEV R}\;,
\eeq
which can account for the large $\n_\m -\n_\t$ mixing angle. This form of the mass matrix is compatible with small and large mixing angle solutions of the solar neutrino problem{\footnote{ In ref.\cite{vis98} it is claimed that for the value of $\e$ in eq.~(6) the large mixing angle solution is favoured.}}. 
The Yukawa matrices which yield the Dirac masses of neutrinos and charged 
leptons have the general structure,
\beq\label{dir}
h_e  \sim\ \e^a\ \left(\begin{array}{ccc}
    \e^3 & \e^2 & \e^2 \\[1ex]
    \e^2 & \e   & \e   \\[1ex]
    \e   & 1    & 1
    \end{array}\right) \;, \quad
h_{\n}  \sim\ \e^a\ \left(\begin{array}{ccc}
    \e^{3-a} & \e^{2-a} & \e^{2-a} \\[1ex]
    \e^{2-a} & \e^{1-a} & \e^{1-a} \\[1ex]
    \e       & 1        & 1
    \end{array}\right) \;.
\eeq
The Yukawa matrix for the right-handed neutrinos can always be chosen
diagonal,
\beq
h_{r}  \sim\ \left(\begin{array}{ccc}
    \e^{4-2a} & 0 & 0 \\[1ex]
    0 & \e^{2-2a} & 0 \\[1ex]
    0  & 0  & 1
    \end{array}\right) \;.
\eeq
The corresponding unitary transformation does not change the structure of $h_{\n}$.
In eqs. (7)-(9) factors $\co(1)$ have been omitted and we assume that there is no degeneracy in the right-handed neutrino mass matrix.

In models with gravity mediated supersymmetry breaking one usually assumes
universal soft breaking terms at the GUT scale,
\beq\label{universal}
{\wt m_l}^2 = {\wt m_e}^2 = M^2 \one \;, \quad A_e = h_e A\;,
\quad A_{\n} = h_{\n} A\;.
\eeq
Renormalization effects change these matrices significantly at lower scales.
As a consequence the flavour structure in the scalar sector is different
from the one in the fermionic sector. Integrating the renormalization group
equations from the GUT scale, and taking the decoupling of the heavy neutrinos
at their respective masses $M_k$ into account, one obtains at scales
$\m \ll M_1$,
\bea\label{softrge}
\d\wt m^2_{l ij} &\simeq& -{1\over 8\pi^2}(3 M^2 + A^2)
                      h^\dg_{\n ik}\ln{\L_{GUT}\over M_k} h_{\n kj}\;,\NO\\ 
\d A_{e ij} &\simeq&  -{1\over 8\pi^2} A
                      (h_eh^\dg_\n)_{ik}\ln{\L_{GUT}\over M_k} h_{\n kj}\;.
\eea
In the following we shall discuss decay rates for lepton number changing
radiative transitions to leading order in $\e$ and we will not be able to
discuss factors $\co(1)$. We therefore neglect terms $\sim \ln{\e^2}$ which
reflect the splitting between the heavy neutrino masses and evaluate
$\ln(\L_{GUT}/M_k)$ for an average mass $\Bar M = 10^{12}$~GeV. 
In eqs.~(\ref{softrge}) this yields the overall factor
$\ln(\L_{GUT}/ \Bar M) \sim 10$. The flavour structure of the left-left
scalar mass matrix is then identical to the one of the neutrino mass matrix,
\beq\label{smix2}
\d\wt m^2_{l ij} \sim {1\over 8\pi^2}(3 M^2 + A^2)
                      \ln{\L_{GUT}\over \Bar M}\ \e^{2a}\
                      \left(\begin{array}{ccc}
                      \e^2  & \e  & \e \\[1ex]
                      \e  &  1  & 1 \\[1ex]
                      \e  &  1  & 1 
                      \end{array}\right)\;.
\eeq
For the flavour changing left-right scalar mass matrix one obtains
\beq\label{smix3}
v_1 \d A_{e ij} \sim  {1\over 8\pi^2} A m_\t
                   \ln{\L_{GUT}\over \Bar M}\ \e^{2a}\
                      \left(\begin{array}{ccc}
                      \e^3  & \e^2  & \e^2 \\[1ex]
                      \e^2  & \e    & \e   \\[1ex]
                      \e    &  1    & 1 
                      \end{array}\right)\;.
\eeq

For a wide class of supergravity models the possibilities of supersymmetry
breaking can be parametrized by vacuum expectation values of moduli fields
$T_\a$ and the dilaton field $S$ \cite{bri94}. The structure of the soft 
breaking terms is determined by the modular weights of the various superfields.
An interesting structure arises if the theory possesses both, modular 
invariance and a chiral U(1) symmetry. In this case the supersymmetry 
breaking scalar mass terms are directly related to the charges of the 
corresponding superfields
\cite{dud96},
\beq\label{modular}
{\wt m}^2_{ij} = \left((1 + B_i(\Q_\a))\d_{ij} + 
                 |Q_i-Q_j| C_{ij}(\Q_{\a}) \times \e^{|Q_i-Q_j|}\right) M^2\;,
\eeq
where the $\Q_\a$ parametrize the direction of the goldstino in the moduli
space. For pure dilaton breaking, $\Q_\a = 0$, one has $C_{ij} = 0$ and
the soft breaking terms are flavour diagonal. In the general case, instead, we get from eq.~(\ref{modular}), 
\beq\label{softmod}
{\wt m}^2_l  \sim\ \left(\begin{array}{ccc}
    1  & \e  & \e \\[1ex]
    \e & 1   & 0  \\[1ex]
    \e & 0   & 1
    \end{array}\right)\ M^2 \;, \quad
{\wt m}^2_e  \sim\ \left(\begin{array}{ccc}
    1    & \e & \e^2 \\[1ex]
    \e   & 1  & \e     \\[1ex]
    \e^2 & \e & 1
    \end{array}\right)\ M^2 \;.
\eeq
Note, that the zeros in ${\wt m}^2_l$ occur since the lepton doublets of
the second and the third family carry the same U(1)$_F$ charge. 
The trilinear soft breaking terms are also affected by modular invariance
\cite{dud96}. The effect is to increase the branching ratio of lepton flavour violating processes. In order to obtain  
 a lower bound, we shall take the trilinear soft breaking
 terms flavour diagonal  and we shall only consider the effects of the lepton flavour changing scalar 
mass terms in the modular invariance case.

The scalar mass matrices (\ref{smix2}), (\ref{smix3}) and (\ref{softmod})
are given in the weak eigenstate basis. In order to discuss the radiative
transitions $\m \rightarrow e \g$ and $\t \rightarrow \m \g$ we have to
 change to a mass eigenstate basis of the charged leptons. 
The Yukawa matrix $h_e$ can be diagonalized by a bi-unitary transformation,
$U^\dg h_e V = h_e^D$. To leading order in $\e$ the matrices $U$ and $V$ are
given by
\beq
U = \left(\begin{array}{ccc}
    1      & a\e  & b\e^2 \\[1ex]
    -a\e   & 1    & f\e   \\[1ex]
    -b\e^2 & -f\e & 1
    \end{array}\right)\;, \quad
V = \left(\begin{array}{ccc}
    1     & (ca'-sb')\e & (sa'+cb')\e \\[1ex]
    -a'\e & c           & s           \\[1ex]
    -b'\e & -s          & c
    \end{array}\right)\;,
\eeq
where $c=\cos{\vf}$ and $s=\sin{\vf}$; $a,b,a'$ and $b'$ depend on the coefficients  $\co(1)$ in $h_e$ which are not given in eq.~(\ref{dir}). The scalar mass matrices transform
as $V^\dg \d \wt m_l^2 V$, $U^\dg v_1 A_e V$, $V^\dg \wt m^2_l V$,
$U^\dg \wt m_e^2 U$. One easily verifies that the form of the matrices 
given in eqs.~(\ref{smix2}), (\ref{smix3}) and (\ref{softmod}) is invariant
under this transformation.
 
Given the Yukawa matrices and the scalar mass matrices it is
straightforward to calculate the rates for radiative transitions. The 
transition $\m \rightarrow e \g$  has the form
\beq\label{transamp}
\cm_\m = ie \bar{u}_e(p-q)\s_{\m\n} q^{\n} 
(A_{L12} P_L + A_{R12} P_R) u_{\m}(p)\;,
\eeq
where $P_L$ and $P_R$ are the projectors on states with left- and
right-handed chirality, respectively. The corresponding branching ratio
is given by
\beq\label{brmu}
BR(\m \rightarrow e\g) = 384\p^3 \a {v^4\over m_\m^2}
  (|A_{L12}|^2 + |A_{R12}|^2)\;,
\eeq
where $v=(8G_F^2)^{1/4}\simeq 174$~GeV is the Higgs vacuum expectation value.

At one-loop order the transition amplitudes for a left(right)-handed muon 
$A_{L(R)}^{(b)}$ and $A_R^{(w)}$  involve neutral and
charged gauginos, respectively. Note, that the amplitude $A_L^{(w)}$ is
suppressed by inverse powers of the heavy neutrino masses $M_i$. 
The radiative transition changes chirality. Amplitudes, where the
chirality change is due to the gaugino require a left-right scalar 
transition and one or two scalar flavour changes (figs.~(1.a)-(1.d)).

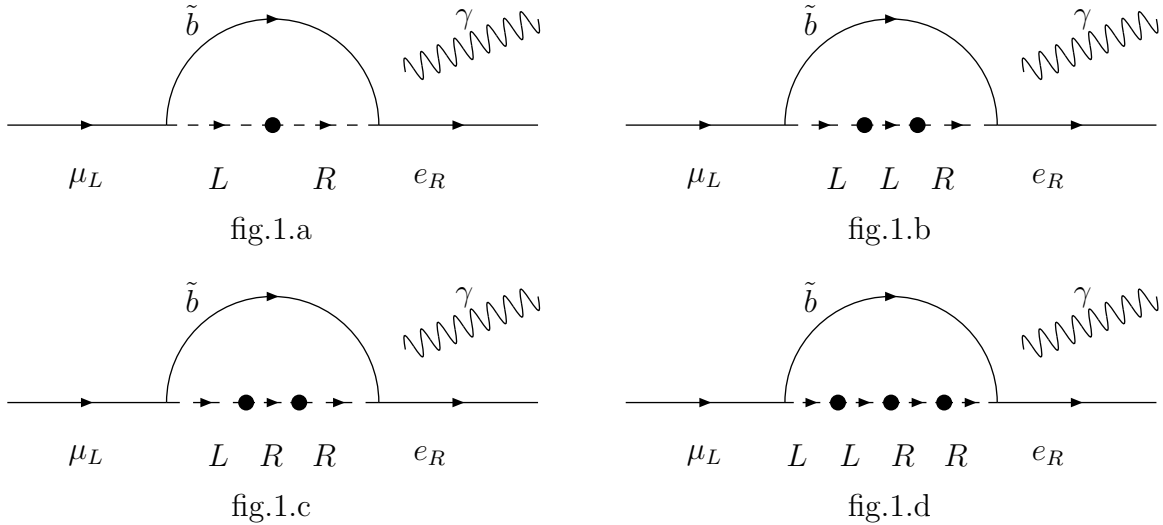
\begin{picture}(200,100)(65,0)
\ArrowArcn(150,50)(40,180,360)
\ArrowLine(50,50)(110,50)
\ArrowLine(190,50)(250,50)
\DashArrowLine(110,50)(150,50){5}
\Vertex(150,50){3}
\DashArrowLine(150,50)(190,50){5}
\Photon(200,70)(250,90){5}{8}
{\Text(80,30)[]{$\mu_L$}}
{\Text(210,30)[]{$e_R$}}
{\Text(130,30)[]{$L$}}
{\Text(170,30)[]{$R$}}
{\Text(120,90)[]{$\tilde{b}$}}
{\Text(225,90)[]{$\gamma$ }}
\Text(150,10)[]{fig.1.a}
\end{picture}
\begin{picture}(200,100)(35,0)
\ArrowLine(50,50)(110,50)
\ArrowLine(190,50)(250,50)
\DashArrowLine(110,50)(140,50){5}
\ArrowArcn(150,50)(40,180,360)
\Vertex(140,50){3}
\Text(130,30)[]{$L$}
\DashArrowLine(140,50)(160,50){5}
\Vertex(160,50){3}
\Text(150,30)[]{$L$}
\DashArrowLine(160,50)(190,50){5}
\Text(170,30)[]{$R$}
\Photon(200,70)(250,90){5}{8}
{\Text(80,30)[]{$\mu_L$}}
{\Text(210,30)[]{$e_R$}}
{\Text(120,90)[]{$\tilde{b}$}}
{\Text(225,90)[]{$\gamma$ }}
\Text(150,10)[]{fig.1.b}
\end{picture}

\begin{picture}(200,100)(65,0)
\ArrowArcn(150,50)(40,180,360)
\ArrowLine(50,50)(110,50)
\ArrowLine(190,50)(250,50)
\DashArrowLine(110,50)(140,50){5}
\Vertex(140,50){3}
\Text(130,30)[]{$L$}
\DashArrowLine(140,50)(160,50){5}
\Vertex(160,50){3}
\Text(150,30)[]{$R$}
\DashArrowLine(160,50)(190,50){5}
\Text(170,30)[]{$R$}
\Photon(200,70)(250,90){5}{8}
{\Text(80,30)[]{$\mu_L$}}
{\Text(210,30)[]{$e_R$}}
{\Text(120,90)[]{$\tilde{b}$}}
{\Text(225,90)[]{$\gamma$ }}
\Text(150,10)[]{fig.1.c}
\end{picture}
\begin{picture}(200,100)(35,0)
\ArrowArcn(150,50)(40,180,360)
\ArrowLine(50,50)(110,50)
\ArrowLine(190,50)(250,50)
\DashArrowLine(110,50)(130,50){5}
\Vertex(130,50){3}
\Text(115,30)[]{$L$}
\DashArrowLine(130,50)(150,50){5}
\Vertex(150,50){3}
\Text(135,30)[]{$L$}
\DashArrowLine(150,50)(170,50){5}
\Vertex(170,50){3}
\Text(155,30)[]{$R$}
\DashArrowLine(170,50)(190,50){5}
\Text(175,30)[]{$R$}
\Photon(200,70)(250,90){5}{8}
{\Text(80,30)[]{$\mu_L$}}
{\Text(210,30)[]{$e_R$}}
{\Text(120,90)[]{$\tilde{b}$}}
{\Text(225,90)[]{$\gamma$ }}
\Text(150,10)[]{fig.1.d}
\end{picture}

From eq.~(\ref{scalarm}) and (\ref{universal}) one reads off,
\beq\label{lrmix}
{\wt M_{RL}}^2 = (A + \m \tan{\b}) M_e \;,
\eeq
where $\tan{\b} = v_2/v_1$, and $M_e=h_e v_1$ is the charged lepton mass 
matrix. Amplitudes with chirality change of the external muon have one scalar
flavour change to leading order in $\e$ for neutral (fig.~(1.e)) and charged 
(fig.~(1.f)) gaugino.

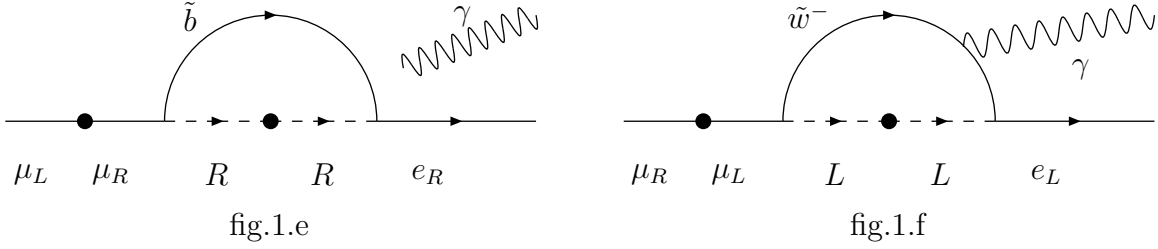
\begin{picture}(200,100)(65,0)
\ArrowArcn(150,50)(40,180,360)
\ArrowLine(50,50)(110,50)
\Vertex(80,50){3}
\Text(60,30)[]{$\mu_L$}
\Text(90,30)[]{$\mu_R$}
\ArrowLine(190,50)(250,50)
\DashArrowLine(110,50)(150,50){5}
\Vertex(150,50){3}
\DashArrowLine(150,50)(190,50){5}
\Photon(200,70)(250,90){5}{8}
{\Text(210,30)[]{$e_R$}}
{\Text(130,30)[]{$R$}}
{\Text(170,30)[]{$R$}}
{\Text(120,90)[]{$\tilde{b}$}}
{\Text(225,90)[]{$\gamma$ }}
\Text(150,10)[]{fig.1.e}
\end{picture}
\begin{picture}(200,100)(35,0)
\ArrowArcn(150,50)(40,180,360)
\ArrowLine(50,50)(110,50)
\Vertex(80,50){3}
\Text(60,30)[]{$\mu_R$}
\Text(90,30)[]{$\mu_L$}
\ArrowLine(190,50)(250,50)
\DashArrowLine(110,50)(150,50){5}
\Vertex(150,50){3}
\DashArrowLine(150,50)(190,50){5}
\Photon(178,78)(250,90){5}{8}
{\Text(210,30)[]{$e_L$}}
{\Text(130,30)[]{$L$}}
{\Text(170,30)[]{$L$}}
{\Text(120,90)[]{$\tilde{w}^{-}$}}
{\Text(225,70)[]{$\gamma$ }}
\Text(150,10)[]{fig.1.f}
\end{picture}

Simple compact expressions can be given for the various transition
amplitudes if one expands the scalar mass matrices around the dominant
universal mass matrix $M \one $, i.e., $\wt m^2_l = M^2 \one + \d\wt m^2_l$,
$\wt M^2_{L,R} = M^2 \one + \d\wt M^2_{L,R}$. For the bino ($b$) and chargino ($w^-$) 
contributions to the transition amplitude one obtains,
\beq\label{ampli} 
A^{(b)}_{L,R} = {\a\over 4\pi \cos^2{\Q_W}} \bar A^{(b)}_{L,R}\;, \quad
A^{(w^-)}_R = {\a\over 8\pi \sin^2{\Q_W}} \bar A^{(w)}_{R}\;, 
\eeq
where
\bea
\bar A^{(b)}_{L12} &=& Y_L Y_R m_b \left((\wt M_{RL}^2)_{12}{\pa\over \pa M^2} 
+{1\over 2}(\d\wt M_R^2 \wt M_{RL}^2 + \wt M_{RL}^2\d\wt M_L^2)_{12} 
{\pa^2\over \pa (M^2)^2}\right.\NO\\
&&\left.\hspace{2cm} + {1\over 3!}(\d\wt M_R^2 \wt M_{RL}^2 \d\wt M_L^2)_{12}
   {\pa^3\over \pa (M^2)^3}\right)F(m_b^2,M^2)\NO\\
&&-Y_R^2 m_{\m}(\d\wt M_R^2)_{12}{\pa\over \pa M^2} G(m_b^2,M^2)+...\;,
\label{bino}\\
\bar A^{(w)}_{R12} &=& -m_{\m}(\d\wt m_l^2)_{12}
             {\pa\over \pa M^2} G(M^2,m_w^2)+...\;. \label{wino}
\eea
Here $m_b$, $m_w$ and $M$ are bino, charged wino and scalar masses, and $Y_L = -1/2$ 
and $Y_R = -1$ are the hypercharges of the lepton multiplets $l_L$ and $e_R$,
respectively. The amplitude $A^{(b)}_R$ is obtained from $A^{(b)}_L$ by 
interchanging all subscripts $L$ and $R$. In eqs.~(\ref{bino}) and 
(\ref{wino}) the dependence on the gaugino masses and the average scalar mass
$M$ has been separated from the dependence on the lepton flavour beaking 
parameters. The functions $F(m^2,M^2)$ and $G(m^2,M^2)$ read
\bea
F(m^2,M^2) &=& {M^4-m^4+2 m^2 M^2 \ln{m^2\over M^2}\over (M^2-m^2)^3}\;,\\
G(m^2,M^2) &=& {M^6-6 m^2 M^4 + 3 m^4 M^2 + 2 m^6 - 6 m^4 M^2 
               \ln{m^2\over M^2}\over 6 (m^2 - M^2)^4}\;. 
\eea
The mixing between Higgsino and gaugino gives also a contribution to the leading order in $\e$. The diagrams contributing to the amplitude are illustrated in fig.(1.g) and fig.(1.h)

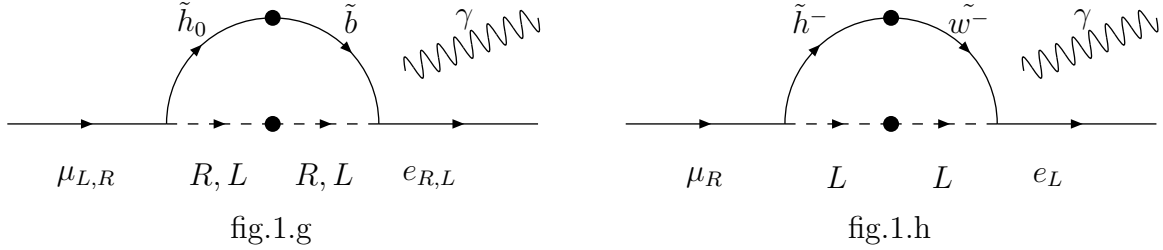
\begin{picture}(200,100)(65,0)
\ArrowArcn(150,50)(40,180,90)
\ArrowArcn(150,50)(40,90,0)
\Vertex(150,90){3}
\ArrowLine(50,50)(110,50)
\ArrowLine(190,50)(250,50)
\DashArrowLine(110,50)(150,50){5}
\Vertex(150,50){3}
\DashArrowLine(150,50)(190,50){5}
\Photon(200,70)(250,90){5}{8}
{\Text(80,30)[]{$\mu_{L,R}$}}
{\Text(210,30)[]{$e_{R,L}$}}
{\Text(130,30)[]{$R,L$}}
{\Text(170,30)[]{$R,L$}}
{\Text(180,90)[]{$\tilde{b}$}}
{\Text(225,90)[]{$\gamma$ }}
\Text(120,90)[]{$\tilde{h}_0$}
\Text(150,10)[]{fig.1.g}
\end{picture}
\begin{picture}(200,100)(35,0)
\ArrowArcn(150,50)(40,180,90)
\ArrowArcn(150,50)(40,90,0)
\Vertex(150,90){3}
\ArrowLine(50,50)(110,50)
\ArrowLine(190,50)(250,50)
\DashArrowLine(110,50)(150,50){5}
\Vertex(150,50){3}
\DashArrowLine(150,50)(190,50){5}
\Photon(200,70)(250,90){5}{8}
{\Text(80,30)[]{$\mu_{R}$}}
{\Text(210,30)[]{$e_{L}$}}
{\Text(130,30)[]{$L$}}
{\Text(170,30)[]{$L$}}
{\Text(180,90)[]{$\tilde{w^-}$}}
{\Text(225,90)[]{$\gamma$ }}
\Text(120,90)[]{$\tilde{h}^{-}$}
\Text(150,10)[]{fig.1.h}
\end{picture}

\noindent and the corresponding amplitudes are given by
\beq\label{mixing1} 
A^{(b,\tilde{h}_{0})}_{L,R} = {\a\over 4\pi \cos^2{\Q_W}} \bar A^{(b,\tilde{h}_{0})}_{L,R}\;, \quad
A^{(w^-,\tilde{h}^{-})}_R = {\a\over 8\pi \sin^2{\Q_W}} \bar A^{(w^-,\tilde{h}^{-})}_{R}\;, 
\eeq
where 
\bea
\bar A^{(b,\tilde{h}_{0})}_{R,L12}&=&Y_{L,R} m_{\m}(\d\wt M_{L,R}^2)_{12} \left\{ \frac{\m^2+\tan \b \m m_b}{m_b^2-\m^2}  {\pa\over \pa M^2} F(\m^2,M^2) + (\m \leftrightarrow m_b) \right\}, \label{binohiggsino}
\eea
\bea
\bar A^{(w^-,\tilde{h}^{-})}_{R12}&=& - m_{\m}(\d\wt m_l^2)_{12}  \left\{ \frac{\m^2+\tan \b \m m_w}{m_w^2-\m^2} {\pa\over \pa M^2} H(\m^2,M^2)+ (\m 
\leftrightarrow m_w) \right\},\label{winohiggsino}
\eea
with 
\beq
H(m^2,M^2)=\frac{3 M^4-4 m^2 M^2 + m^4+2 M^4 \log\frac{m^2}{M^2}}{(m^2-M^2)^3}.
\eeq
These expressions are correct at leading order in $v/M_{SUSY}$.
It is clear from eqs.~(\ref{winohiggsino}) and (\ref{binohiggsino}), that for $\mu \gg m_{b,w}$ the Higgsino-gaugino mixing contributions are suppressed compared to the diagrams with just a gaugino exchange. So for the discussion of the uncertainties on the $BR( \m \rightarrow e \g)$ and $BR(\t \rightarrow \m \g)$ in the three classes of models, the lowest bound on these branching ratios are obtained when the higgsino-gaugino mixing is neglected. For the upper bound, the dominant contribution is coming from the left-right scalar transition.

From eqs.~(\ref{softmod}), (\ref{lrmix}) (\ref{bino}) and (\ref{wino}) one 
easily obtains the transition amplitudes for the models with modular invariance
to leading order in $\e$,
\beq
\bar A^{(b)}_{L12} 
=m_\m \e \left\{  {1\over 4} m_b\ (A + \m \tan{\b}) 
 M^2 {\pa^2\over \pa (M^2)^2}F(m_b^2,M^2) 
- M^2 {\pa\over \pa M^2} G(m_b^2,M^2) \right\} ,\label{modbino}\\
\eeq
\beq
\bar A^{(w)}_{R12}
=-m_{\m}\ \e\ M^2 {\pa\over \pa M^2} G(M^2,m_w^2)\; .\label{modwino}
\eeq
The corresponding amplitudes for models with radiatively induced lepton 
flavour mixing are obtained from eqs.~(\ref{smix2}), (\ref{smix3}), 
(\ref{bino}) and (\ref{wino}),
\bea\label{amprad}
\bar A^{(b)}_{L12} 
&=& {1\over 2} m_b\ v_1 (A_e^\dg)_{12}
    {\pa\over \pa (M^2)} F(m_b^2,M^2) \NO\\
&\simeq& {1\over 16 \pi^2} m_b A m_\m \e^{2a+1} \ln{\L_{GUT}\over \Bar M}
    {\pa\over \pa (M^2)} F(m_b^2,M^2)\;,\label{radbino}\\
\bar A^{(w)}_{R12}
&=&-m_{\m}\ ({\wt m^2_l})_{12} {\pa\over \pa M^2} G(M^2,m_w^2) \NO\\
&\simeq& -{1\over 8\pi^2}(3 M^2 + A^2) m_\m \e^{2a+1} 
    \ln{\L_{GUT}\over \Bar M}
    {\pa\over \pa (M^2)} G(M^2,m_w^2)\;.\label{radwino}
\eea
Here we have used $m_\m \simeq \e m_\t$ in eq.~(\ref{radbino}).    

Based on the results for $\m \rightarrow e \g$ one can immediately write down 
the rate for the process $\t \rightarrow \m \g$. Using $\G_\t \simeq 5 
(m_\t/m_\m)^5 \G_\m$, one obtains for the branching ratio,
\beq
BR(\t \rightarrow \m\g) \simeq {384\p^3\over 5} \a {v^4\over m_\t^2}
  (|A_{L23}|^2 + |A_{R23}|^2)\;.
\eeq
The amplitudes $\bar A^{(b)}_{L23}$ and $\bar A^{(w)}_{R23}$ are easily
obtained from eqs.~(\ref{modbino}) - (\ref{radwino}). For models with
modular invariance one has
\beq
\bar A^{(b)}_{L23} \sim {m_\t\over m_\m} \bar A^{(b)}_{L12}\;,\quad
\bar A^{(w)}_{L23} = 0\;.
\eeq
Note, that the vanishing of $\bar A^{(w)}_{L23}$ is a direct consequence
of the fact that the lepton doublets of the second and third generation
have the same chiral charge. In models with radiatively induced flavour
change one obtains
\beq
\bar A^{(b)}_{L23} \sim {m_\t\over m_\m\e} \bar A^{(b)}_{L12}\;,\quad
\bar A^{(w)}_{L23} \sim {m_\t\over m_\m\e} \bar A^{(w)}_{L12} \;.
\eeq

The branching ratios for $\m \rightarrow e \g$ and $\t \rightarrow \m \g$
strongly depend on the gaugino and scalar masses. Collecting all factors
in eqs.~(\ref{brmu}), (\ref{ampli}) and (\ref{modbino}) - (\ref{radwino})
one finds for the order of magnitude of the branching ratios in the case
$m_b \sim m_w \sim M \sim A \sim v$ for the models with modular invariance (MI)
and radiatively induced flavour violation (RI), respectively,
\bea
BR_{(MI)}(\m \rightarrow e \g) &\sim& \a^3\e^2
\sim 5 BR_{(MI)}(\t \rightarrow \mu \g)\;, \label{modord}\\
BR_{(RI)}(\m \rightarrow e \g) &\sim& 0.1\ \a^3\e^{4a+2}
\sim 5\e^2 BR_{(RI)}(\t \rightarrow \mu \g)\;. \label{radord}
\eea
For large Yukawa couplings, i.e. $a=0$, the branching ratio
$BR(\m \rightarrow e \g)$ is of the same order in $\e^2$ for both classes of 
models. The numerical factor in eq.~(\ref{radord}) occurs, because the flavour
mixing only arises at one-loop order. The suppression is not stronger since 
the one-loop contribution is enhanced by a large logarithm,  
$\ln{\L_{GUT}/\Bar M}$. With $\e^2\sim 1/300$, one obtains 
$BR_{(MI)}(\m \rightarrow e \g) \sim 10^{-9}$, more than one order of 
magnitude above the experimental upper limit. In models with modular
invariance the branching ratios for $\m \rightarrow e \g$ and
$\t \rightarrow \mu \g$ are of the same order in $\e^2$. In the case of
radiatively induced flavour violation $BR_{(RI)}(\t \rightarrow \mu \g)$
is enhanced by $1/\e^2$ due to the large mixing between leptons of the
second and third generation. 

\setcounter{figure}{1}
\begin{figure}[tbh]
\begin{center}
\mbox{\epsfxsize=14.cm\epsfysize=8.cm\epsffile{BRmu4.eps}}
\end{center}
\par
\caption{{\it Predicted range for $BR(\mu \rightarrow e \gamma)$ as  function of
 the gaugino mass in the three cases (see text): modular invariance (gray lines)
, radiatively induced flavour violation with large Yukawa couplings (dashed lines)
 and small Yukawa couplings (black lines). The straight line correspond to the
 experimental bound on $BR(\m \rightarrow e \g)< 4.9~\cdot~10^{-11}$}\cite{parti
cles}.    }
\end{figure}

In order to determine the uncertainty of the theoretical predictions one
has to vary the various supersymmetry breaking parameters in a range
consistent with present experimental limits. For gaugino masses and the
average scalar mass our choice is $m_b = 100\ldots 500$~GeV,
$m_w = 100\ldots 500$~GeV, $M= 100\ldots 500$~GeV, $A=0\ldots M$,
$A + \m \tan{\b} = 0\ldots M$. We know the transition amplitude
only up to a factor $\co(1)$. We therefore also neglect neutralino and chargino
mixings and we assumed for simplicity that the gauginos masses are equal. To estimate these uncertainties we increase the upper bound on the branching ratio
by a factor of 5 and decrease the lower bound by a factor $1/5$. The
result for $BR(\m \rightarrow e \g)$ is shown in fig.~2 as function of
the gaugino mass. The upper bound is given by the bino contribution with
large mixing between `left' and `right' scalars ($M = 100$~GeV, 
$A = M$, $A + \m \tan{\b} = M$); the lower limit is determined by the
chargino contribution ($M = 500$~GeV, $A = 0$, $A + \m \tan{\b} = 0$).
For $\t \rightarrow \mu \g$ the predicted branching ratio lies below the
present upper experimental bound in all cases (cf. fig.~(3)).

\begin{figure}[tbh]
\begin{center}
\mbox{\epsfxsize=14.cm\epsfysize=8.cm\epsffile{BRtau4.eps}}
\end{center}
\par
\caption{{\it Predicted range for $BR(\tau \rightarrow  \mu  \gamma)$ as  function of
 the gaugino mass in the three cases (see text): modular invariance (gray lines)
, radiatively induced flavour violation with large Yukawa couplings (dashed lines)
 and small Yukawa couplings (black lines). The straight line correspond to the experimental bound on $BR(\t \rightarrow \m \gamma)<3~\cdot~10^{-6}$}\cite{particles}.} 
\end{figure}

For most of the parameter space the prediction for $BR(\m \rightarrow e \g)$
in  models with modular invariance exceeds the experimental upper limit. 
Hence, this pattern of supersymmetry breaking appears to be disfavoured.
For radiatively induced flavour violation and large Yukawa couplings the
predicted range of branching ratios includes the present upper limit. An
improvement of the sensitivity by two orders of magnitude would cover the 
entire parameter space. In the case of small Yukawa couplings the branching 
ratio is suppressed by $\e^4 \sim 10^{-5}$, and therefore far below the 
experimental limit.

For $\t \rightarrow \mu \g$ the largest branching ratio is obtained for
radiatively induced flavour mixing with large Yukawa couplings. This is a direct consequence of the large mixing between neutrinos of the second and the third generation. The observation of this radiative transition would therefore be of great significance.

\end{document}